\newcommand\rd{\mathrm d}
\documentclass[%
 reprint,
 amsmath,amssymb,
 aps,
]{revtex4-2}

\usepackage{graphicx}
\usepackage{dcolumn}
\usepackage{bm}
\usepackage{amsmath}
\usepackage{svg}
\usepackage{caption}
\usepackage{todonotes}
\usepackage[justification=centering]{caption}

\begin{document}

\preprint{APS/123-QED}

\title{Monte Carlo modelling of the linear Breit-Wheeler process within the GEANT4 framework}

\author{R. A. Watt}
\author{S. J. Rose}
\author{B. Kettle}
\author{S. P. D. Mangles}

\affiliation{%
 The John Adams Institute for Accelerator Science, Imperial College London, London, SW7 2AZ, United Kingdom}%

\date{February 9, 2023}

\begin{abstract}
A linear Breit-Wheeler module for the code Geant4 has been developed. This allows signal-to-noise ratio calculations of linear Breit-Wheeler detection experiments to be performed within a single framework.
The interaction between two photon sources is modelled by treating one as a static field, then photons from the second source are sampled and tracked through the field.
To increase the efficiency of the module, we have used a Gaussian process regression, which can lead to an increase in the calculation rate by a factor of up to 1000.

To demonstrate the capabilities of this module, we use it to perform a parameter scan, modelling an experiment based on that recently reported by Kettle \textit{et al.} \cite{KettlearXiv2021}.
We show that colliding $50\,$fs duration $\gamma$-rays, produced through bremsstrahlung emission of a $100\,$pC, $2\,$GeV laser wakefield accelerator beam, with a $50\,$ps X-ray field, generated by a germanium burn-through foil heated to temperatures $>\,150\,$eV, this experiment is capable of producing $>1\,$ Breit-Wheeler pair per shot.

\end{abstract}

\maketitle

\section{\label{sec:1} Introduction}
The linear Breit-Wheeler (BW) process is the annihilation of two photons to produce an electron positron pair ($\gamma \gamma \rightarrow e^+ e^-$) and is the simplest mechanism by which matter can be generated from light \cite{Breit1934}. 
The process was first predicted in 1934, however, despite the long time that has passed, the annihilation of real photons has never been directly observed in the laboratory. 
This is due to the difficulty in generating the high energy-density photon sources required to overcome the relatively small cross-section.
However, such photon sources are routinely found in astrophysical environments, and the BW process is predicted to play an important role in a range of phenomena, with examples including gamma ray bursts and the emission from quasars \cite{Piran2005, Bonometto1971}.
Also for photons propagating through the intergalactic medium there is a high energy ($<100\,\mathrm{GeV}$) cut-off in the cosmic gamma ray spectrum observed at Earth due to BW annihilation with the cosmic microwave background \cite{Nikishov1961, Gould1967}.
This mechanism was thought to be well understood. However, recent observations have found a larger number of high energy photons from quasar 3C 279 reaching Earth than expected \cite{Aleksic2011}.
This demonstrates the need for more experimental work, to better understand these astrophysical environments.

Recently, the STAR collaboration \cite{adam2021measurement} have reported the observation of the BW process from quasi-real photons in the peripheral collisions of high energy ions.
However, observing the annihilation of real photons remains elusive.
With the advance of high power laser systems, generating high energy-density photon sources in the laboratory has become possible.
This has led to the proposal of several laser based BW detection experimental schemes utilising current and future facilities.
Pike \textit{et al.} \cite{Pike2014} have suggested a scheme involving the generation of a thermal X-ray field with a temperature of $\sim 300\,$eV, using a laser heated hohlraum.
A beam of GeV gamma rays, produced by bremsstrahlung emission of a laser wakefield accelerated (LWFA) electron beam in a high Z material target, then interacts with the X-ray photons.
To produce the thermal X-ray field requires a facility such as the National Ignition Facility (NIF).
Pike \textit{et al.} predict up to $10^5$ pairs can be produced in a single shot. 
A second scheme involving the interaction between two symmetrical photon sources has been proposed by Ribeyre \textit{et al.} \cite{Ribeyre2016}.
With symmetrical sources, the photon energy required to produces the rest mass of the $e^+ e^-$ pair is reduced to the MeV scale.
Ribeyre \textit{et al.} have compared multiple sources and found synchrotron emission from a highly energetic electron beam in an intense laser field and bremsstrahlung emission to be the most efficient methods.
A pair yield of $\sim 10^4$ per shot is predicted with a relatively modest laser energy of $100\,\mathrm{J}$.
Finally, building on this symmetric setup, Drebot \textit{et al.} \cite{Drebot2017} have investigated a scheme involving the interaction of Compton sources produced by $\sim260\,$MeV electron beams and joule class lasers.
The predicted yield of $\sim 10^{-4}$ pairs per shot is far lower than other experimental schemes, however, due to the low laser energy required, a high repetition rate system can be used.
The BW process would then be detected over many shots.

\begin{figure*}
  \includegraphics[width=\linewidth]{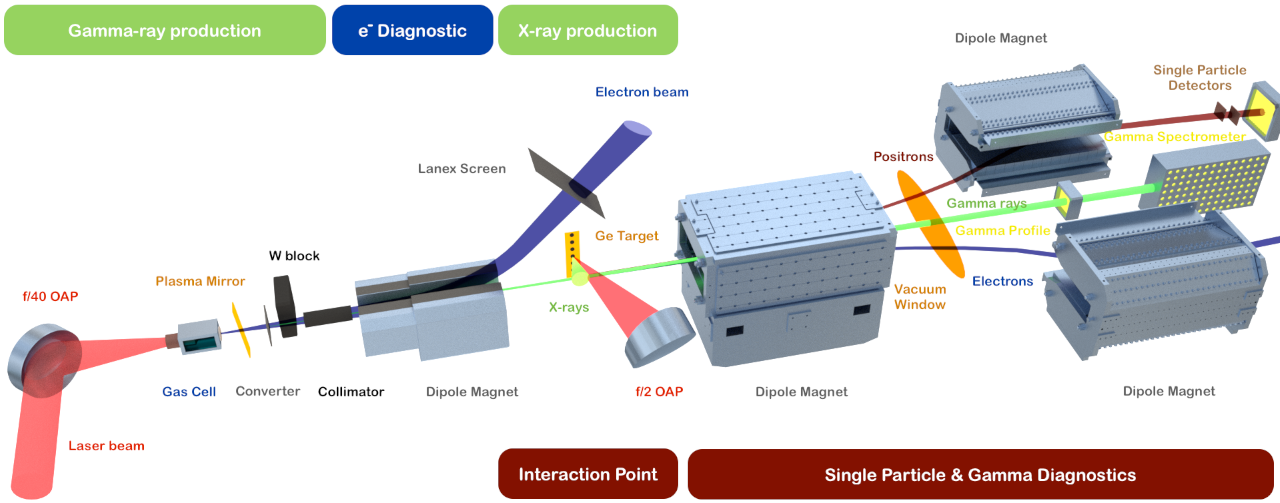}
  \caption{Schematic of Gemini linear BW detection experiment. Starting from the left: a LWFA generates an electron beam which is converted into a gamma ray beam through bremsstrahlung emission in a thin bismuth converter foil. A tungsten collimator and block are placed in the beam path to remove highly divergent gamma rays and those directed towards the X-ray foil respectively. A large number of BH pairs are produced in the converter foil, collimator and block. These are removed with an on-axis magnet before the interaction zone. The gamma rays interact with an X-ray field, generated by a laser heated germanium foil, producing BW pairs. The residual gamma rays continue on axis to a spectrometer. The BW pairs pass through a magnetic chicane to single particle detectors, situated behind lead shielding (not shown). Diagram provided by E. Gerstmayr.}
  \label{fig:experiment}
\end{figure*}

In the Spring of 2018 a BW detection campaign took place at the Gemini laser of the Central Laser Facility at the Rutherford Appleton Laboratory, in the UK. 
The experiment was based on the scheme proposed by Pike \textit{et al.} \cite{Pike2014} and a diagram of the experimental setup is shown in figure \ref{fig:experiment}.
A more in depth review of this experiment can be found in Kettle \textit{et al.} \cite{KettlearXiv2021}.
An electron beam generated by a LWFA interacted with a $1\,$mm bismuth target, producing a $\sim 50\,$fs pulse of gamma rays with a spectrum extending up to 100s of MeV.
This was directed through a $\sim 50\,$ps duration X-ray field, generated by the thermal emission of a laser heated germanium foil which emits strongly between 1$\,$keV and 2$\,$keV \cite{HOARTY2010105}.
Both beams of the Gemini laser are capable of providing 15$\,$J, which is low compared to a high energy facility such as NIF at $2\,$MJ. 
However, the Gemini laser has a much higher repetition rate of 0.05$\,$Hz compared to 1 shot per day.
With a large number of shots a better characterisation of the background can be performed as well as a statistical analysis of the interaction.

All linear BW detection experiments require the generation of high energy ($>0.511\,$MeV) photons.
As a result, the environment is inherently noisy.
There are two main sources of noise: the direct detection of photons by the particle detectors, and photons interacting with the experimental setup, producing background $e^+ e^-$ pairs through the Bethe-Heitler (BH) process \cite{Bethe1934} (the annihilation of a photon with a virtual photon in the nuclear field of an atom).
For a successful experiment, the ratio of signal BW pairs to the background noise should be maximised. 
An estimation of the background noise can be obtained by performing full scale simulations of the passage of high energy particles through the experimental setup. 
Several publicly available Monte Carlo particle tracking codes exist for this purpose, including Geant4, Fluka and MCNP \cite{Geant, Battistoni2015, Briesmeister:2000wra}. 
For the 2018 Gemini campaign Geant4 was used to analyse the background noise \cite{KettlearXiv2021}.
However, the linear BW process is not included within the standard Geant4 physics package.
It is advantageous to perform both signal and noise calculations within a single framework, and the aim of this work has been to develop a linear BW module for Geant4 to enable this.

Geant4 models the passage of individual particles through matter. 
To include the linear BW process within Geant4, it must be modelled within this same framework.
This has been achieved by treating one photon source as a static photon field.
Individual photons from the second source are then sampled and tracked through this field.
In this work, we will refer to a photon from the field as \textbf{\textit{static}} and a photon tracked through the field as \textbf{\textit{dynamic}}.
Using this method, the temporal evolution of the static photon source cannot be accounted for.
This module is therefore suited for modelling experiments with asymmetric photon sources, in which one is constant in time, to a good approximation, over the full interaction.
Figure \ref{fig:experiment} shows an example of such an experiment.

\section{Module Overview}
\label{sec:2}
As discussed in section \ref{sec:1}, this module involves tracking dynamic photons through a static photon field.
The static photon field is represented by a physical volume within the Geant4 computational domain.
It is fully defined by $n(\epsilon, \theta, \phi)$, the number of photons per unit volume with energy between $\epsilon$ and $\epsilon + \mathrm{d}\epsilon$, and travelling at an angle between $\theta + \mathrm{d}\theta$ and $\phi + \mathrm{d}\phi$.
For a dynamic photon which enters a static photon field the calculation of the interaction involves two steps:
\begin{enumerate}
\item Calculating the probability of the interaction occurring.
\item Calculating the dynamics of the interaction.
\end{enumerate}

A dynamic photon enters the static photon field at some arbitrary angle in the simulation frame. 
To perform the calculation of step 1, we transform $n(\epsilon, \theta, \phi)$ into a frame where the z-axis is the direction of propagation of the dynamic photon. We will refer to this frame as the dynamic photon frame. 
For the linear BW process to occur, the total energy in the centre-of-mass (c.m.) frame must be larger than twice the rest mass energy of an electron. This sets the following interaction threshold
\begin{equation} \label{eq:COM}
s = 2\epsilon E (1 - \textrm{cos}\,\theta) > 4 m^2c^4 
\end{equation} 
where $E$ is the energy of the dynamic photon. 
If $n(\epsilon, \theta, \phi) = 0$ for all $\epsilon$, $\theta$ and $\phi$ satisfying equation \ref{eq:COM}, the threshold is not met and the probability of the dynamic photon interacting is 0.
In this case the dynamic photon will propagate through the static field unaffected. 
However, if $n(\epsilon, \theta, \phi) \not= 0$ there is a finite probability of the dynamic photon interacting.
To obtain this probability requires the mean free path of the interaction, $\lambda$, which is obtained through the following \cite{Gould1966}
\begin{equation} \label{eq:MFP}
\begin{split}
\frac{1}{\lambda} = \int_0^{2\pi} \rd \phi \int_{0}^\pi \rd \theta \int_0^{\infty}  \rd \epsilon \,\, \sigma_{\textrm{BW}}(s) \,  n( \phi, \theta,  \epsilon)\, (1 - \cos{\theta})
\end{split}
\end{equation}
where the factor of $(1 - \textrm{cos}(\theta))$ accounts for the relative velocity between photons and $\sigma_{\mathrm{BW}}$ is the polarisation averaged, total BW cross-section \cite{Jauch2009}:
\begin{equation} \label{totalCS}
\sigma_{\textrm{BW}} = \frac{\pi}{2} r_e^2 (1 - \beta^2)\bigg[-2\beta(2 - \beta^2) + (3 - \beta^4) \ln\frac{1+ \beta}{1 - \beta}\bigg]
\end{equation}
where $\beta = \sqrt{(1 - 4s^{-1})}$ and $r_e$ is the classical electron radius.
The probability that a dynamic photon propagates a distance $l$ is 
\begin{equation}
P(l) =  \lambda^{-1} e^{-l / \lambda }.
\end{equation}
For each dynamic photon, a value of $l$ is sampled from this distribution.
If $l$ is longer than the static photon field, the dynamic photon will propagate through unaffected.
If $l$ is shorter, the dynamic photon will propagate $l$ and annihilate.

\begin{figure}
  \includegraphics[width=\linewidth]{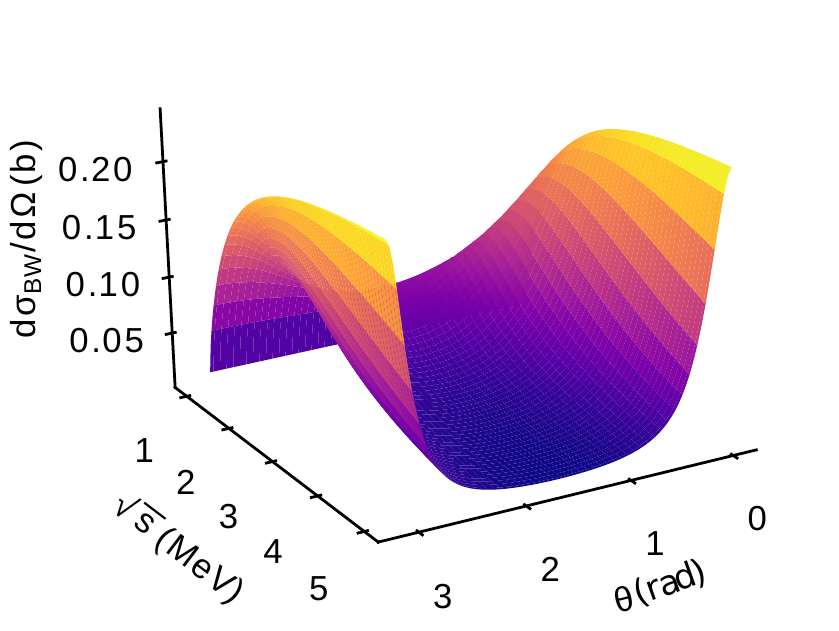}
  \caption{Polarisation averaged BW differential cross-section.}
  \label{fig:diffCS}
\end{figure}

When a dynamic photon  annihilates it is removed from the simulation. We then move onto step 2, determining the dynamics of the interaction. 
First the properties of the annihilating static photon ($\epsilon, \theta, \phi$) must be calculated. These are obtained by sampling from the integrand in equation \ref{eq:MFP}.
In the c.m. frame, both the electron and positron receive an energy, $E_{+,-}$, of $\sqrt{s} / 2$. 
The c.m. frame positron polar scattering angle, $\theta_+$,  is obtained by sampling from the BW differential cross-section \cite{Jauch2009}
\begin{equation} \label{eq:diffCS}
\frac{d\sigma_{\textrm{BW}}}{d\Omega} = \frac{r_e^2\beta}{s}\bigg[\frac{1 + 2\beta \mathrm{sin}^2\theta - \beta^4 - \beta^4 \mathrm{sin}^4\theta}{(1 - \beta^2\mathrm{cos}^2\theta)^2}\bigg]
\end{equation}
which is plotted in figure \ref{fig:diffCS}. 
The positron azimuthal scattering angle, $\phi_+$, is sampled from the uniform distribution $\mathcal{U}(0, 2\pi)$. Using conservation of momentum, the electron polar and azimuthal scattering angles are given by $\theta_- = \pi - \theta_+$ and $\phi_- = \phi_+ - \pi$. From $E_{+,-}$, $\theta_{+,-}$ and $\phi_{+,-}$ the momentum in the c.m. frame, $\textbf{P}_{+,-}$, is calculated. Finally a Lorentz boost is performed into the laboratory frame and an electron and positron are added with the calculated momentum.

\section{Increasing Efficiency with Gaussian Process Regression}
\label{sec:GP}
Due to the low photon density, length of field and value of $\sigma_{\mathrm{BW}}$, the probability that an individual photon will undergo the linear BW process is small. 
To analyse the interaction, many dynamic photons must be simulated. For each dynamic photon $\lambda$ is calculated, involving a triple integral over the static photon variables (equation \ref{eq:MFP}). 
This makes the method outlined in section \ref{sec:2} computationally expensive.

A common solution to this problem is to replace the direct, expensive calculation of $\lambda$ with a lookup table. $\lambda$ is a function of three variables of the dynamic photon, its energy, $E$, and direction, $\theta_d$ and $\phi_d$, through the rotation to the dynamic photon frame. A lookup table would involve discretizing $E$, $\theta_d$ and $\phi_d$ to form a three dimensional grid. At each grid point $\lambda$ is calculated and values are obtained at runtime by interpolating from the grid. However, a lookup table is not appropriate in this case for the following reasons
\begin{enumerate}
    \item $\lambda$ varies rapidly with $E$, $\theta_d$ and $\phi_d$. Therefore, to avoid numerical errors, a fine grid resolution is required. The generation of this lookup table is then computationally expensive.
    \item To avoid numerical errors through extrapolation, the limits of the table must be known prior to the simulation runtime. This can be challenging if the dynamic photons are generated through a multi-step process.
    \item The generation of a uniform dense lookup table does not represent the distribution of the dynamic photon parameters, making it inefficient.
\end{enumerate}

We have used an alternative approach where a Gaussian process regression (GPR) model is trained dynamically and used to quickly calculate $\lambda$ when required. GPR is a Bayesian method for approximating predictions from an expensive physical model. Here, we will only discuss our implementation and not the concepts of GPR. For a full review of GPR refer to Rasmussen and Williams (2006) \cite{Rasmussen2006}.

We can consider equation \ref{eq:MFP} as an expensive function mapping the dynamic photon parameters $\textbf{x} \equiv (E, \theta_d, \phi_d)$ to $\lambda$.
The goal is to replace equation \ref{eq:MFP} with a cheaper function, $f$.
However, $\lambda$ varies rapidly with $\textbf{x}$, therefore, it is more numerically accurate to find a function which returns the natural logarithm of the mean free path estimate
\begin{equation}
    \lambda_a = e^{f(\textbf{x})}
\end{equation}
where $f(\textbf{x})$ is provided by a GPR. 
By using a GPR we have assumed that $f(\textbf{x})$ is Gaussian distributed.
Therefore,  $\lambda_a$ is log-normal distributed with mean and variance given by
\begin{equation} \label{eq:mean-var}
\begin{split}
    \mu_{\lambda_a} &= \mathrm{exp}\Bigg( \mu_f - \frac{\sigma^2_f}{2}\Bigg) \\ \sigma^2_{\lambda_a} &= \Big[\mathrm{exp}(\sigma^2_f) - 1 \Big]\, \mathrm{exp}(2 \mu_f + \sigma_f^2)
\end{split}
\end{equation}
where $\mu_f$ and $\sigma^2_f$ are the mean and variance of $f(\textbf{x})$.

To train the GPR a data set, $\mathcal{D} = \{\mathbf{x}_i, \mathrm{log}(\lambda_i)\}$, is required.
This is generated by calculating equation \ref{eq:MFP} for a limited number of events.
In the units system of Geant4, the dimensions of $\mathbf{x}$ vary over vastly different scales.
Therefore, if the regression model is trained directly on $\mathcal{D}$ it would perform poorly.
To solve this problem $\mathbf{x}$ is normalised using min-max feature scaling.

In this project the open source library libgp \cite{libgp}, was used for the GPR and the implementation can be broken down into the following three stages:
\begin{enumerate}
    \item \textbf{Data accumulation stage}: Equation \ref{eq:MFP} is solved in full for $n_d$ dynamic photons and the result saved, generating $\mathcal{D}$. 
    \item \textbf{Training stage}: After simulating $n_d$ dynamic photons the GPR is trained on $\mathcal{D}$ by optimising the model hyperparameters.
    \item \textbf{Acceleration stage}: For subsequent events, ${\sigma_{\lambda_a}}$ is calculated using equation \ref{eq:mean-var} if ${\sigma_{\lambda_a}} < \sigma_\mathrm{max}$ where $\sigma_\mathrm{max}$ is a user defined limit, $\lambda$ is given by $\mu_{\lambda_a}$. If ${\sigma_{\lambda_a}} > \sigma_\mathrm{max}$ equation \ref{eq:MFP} is solved and the result is appended to $\mathcal{D}$. After another $n_d$ points are added to $\mathcal{D}$ the hyperparameters are again optimised.    
\end{enumerate}
As the simulation progresses $\mathcal{D}$ grows. 
This causes the confidence of the GPR to increase, so it is used more often than the full calculation of equation \ref{eq:MFP}. 
The rate at which dynamic photons are simulated then increases. 
The values of $n_d$ and $\sigma_\mathrm{max}$ can be tuned to optimise the efficiency for an acceptable level of error.

This GPR scheme overcomes all the shortfalls of a lookup table listed above. As the GPR is trained dynamically, no expensive calculations prior to runtime are required. 
Also, no limits on the dynamic photon parameters are set prior to runtime.
This avoids extrapolation as $\lambda$ is calculated in full when the dynamic photon parameters fall well outside the data set. 
Finally, as the GPR is trained on a sample from the dynamic photon population, it is fully representative of the dynamic photon parameters distribution.

\begin{figure}
  \includegraphics[width=\linewidth]{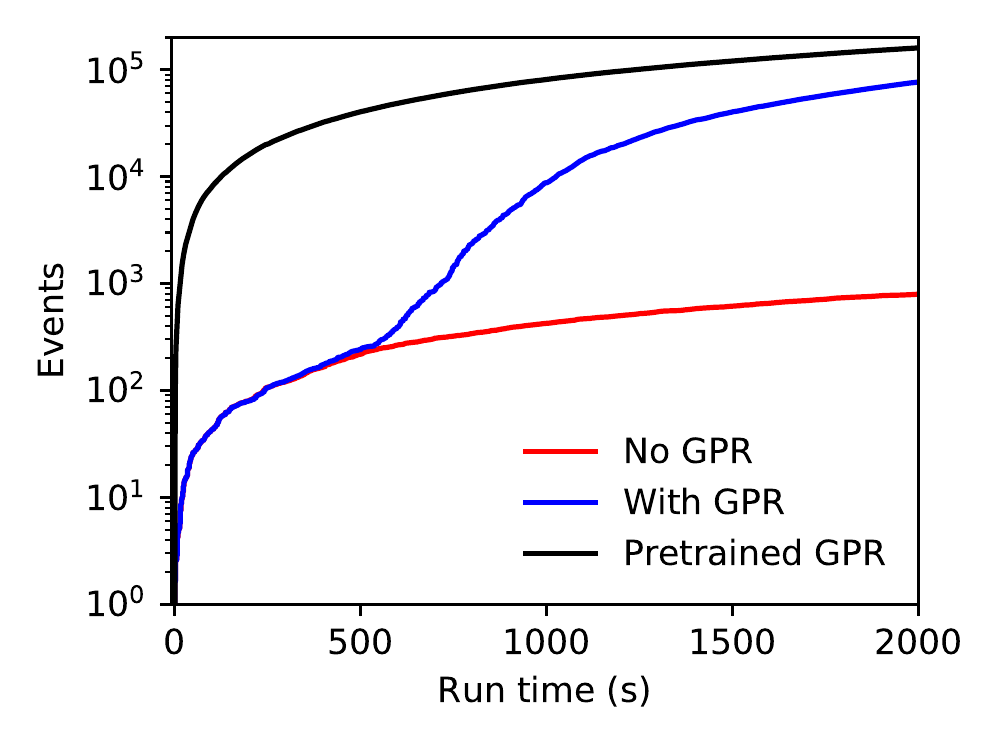}
  \caption{Number of simulated dynamic photons against runtime for the full calculation of equation \ref{eq:MFP} (red), the accelerated GPR scheme (blue) and the pre-trained GPR scheme (black).} 
  \label{fig:GPtime}
\end{figure}

To demonstrate the increase in computational efficiency a test simulation has been carried out for both the full calculation of equation \ref{eq:MFP} and the GPR scheme.
This test simulation involves a gamma ray beam with a divergence of $10\, \mathrm{mrad}$ and a Gaussian energy spectrum with mean $1\, \mathrm{GeV}$ and standard deviation $0.25\, \mathrm{GeV}$, interacting with a $1\,$cm long, $300\,$eV isotropic black body radiation field
\begin{equation}
        n(\epsilon, \theta, \phi) = \frac{\epsilon^2}{\pi^2} \frac{1}{e^{\epsilon / T} - 1} \frac{\sin \theta}{4 \pi}.
\end{equation}
For the GPR scheme the parameters $n_d$ and $\sigma_\mathrm{max}$ are set to 100 and $10^{-2}$ respectively. 
Figure \ref{fig:GPtime} shows the number of dynamic photons simulated against runtime. 
The red curve corresponds to the full calculation of equation \ref{eq:MFP} whereas the blue curve corresponds to the GPR scheme. 
At the start of the simulation figure \ref{fig:GPtime} shows both schemes have the same rate of dynamic photon calculation. 
This is during the data accumulation stage. 
Just after $500\,$s, the GPR enters the training stage. 
Beyond this point the GPR model dynamic photon computation rate increases rapidly. 
After $2000\,$s the GPR scheme has calculated 100 times more dynamic photons.
When performing experimental parameter optimisation, the same static photon field is often used in multiple simulations.
In this case the GPR model can be saved after the end of a simulation and reused in a later one.
This avoids the slow data accumulation stage and provides an even greater increase in efficiency as shown by the black curve in figure \ref{fig:GPtime}.

\begin{figure}
  \includegraphics[width=\linewidth]{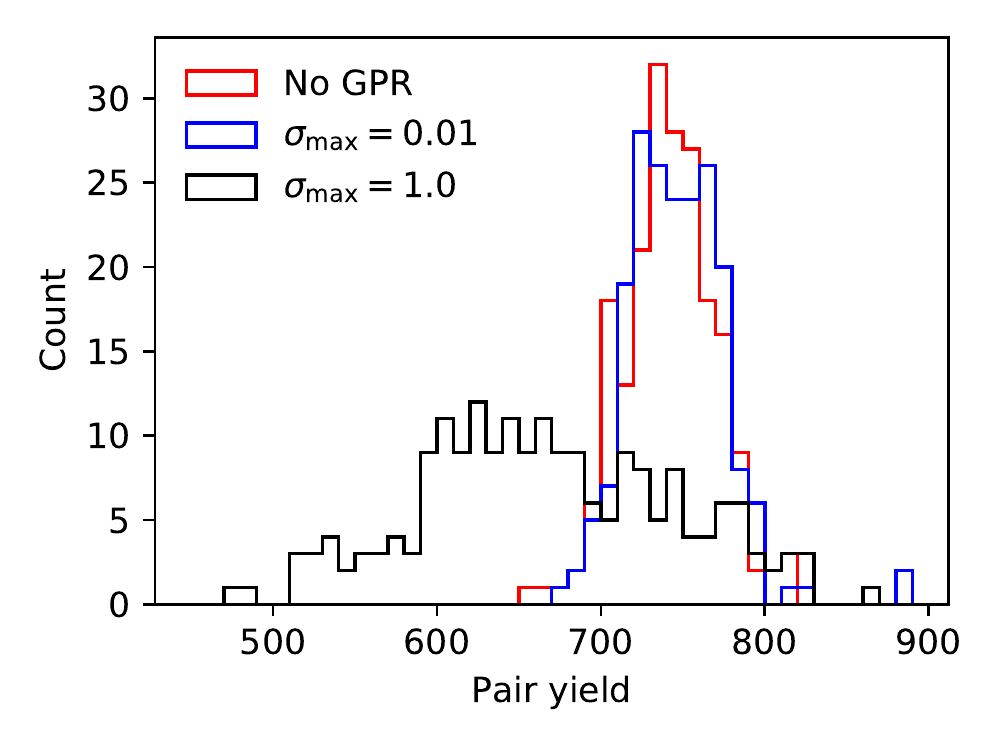}
  \caption{Histograms of number of positrons generated per simulation. The red curve shown the full calculation of equation \ref{eq:MFP} and the blue and black curves show calculations performed with the GPR scheme with a $\sigma_\mathrm{max}$ of $0.01$ and $1.0$ respectively.}
\label{fig:count-hist}
\end{figure}

To test if the GPR scheme is accurately emulating the full calculation of equation \ref{eq:MFP}, simulations can be carried out for both algorithms and the number of pairs generated, $N_\mathrm{BW}$, compared.
However, both algorithms include uncertainty in $N_\mathrm{BW}$ due to the finite number of events simulated.
Therefore, an ensemble of simulations should be carried out and the distributions over $N_{BW}$ compared.
The result of this is shown in figure \ref{fig:count-hist} for two different values of $\sigma_\mathrm{max}$.
When $\sigma_\mathrm{max} = 0.01$ there is little difference between the distributions of the two algorithms, suggesting that the GPR scheme is accurately emulating \ref{eq:MFP}.
However, figure \ref{fig:count-hist} also shows that if $\sigma_\mathrm{max}$ is set too large, this is not the case and the GPR algorithm introduces errors to the estimate.

\section{Module Discussion}
Fully modelling the interaction between two photon sources where both are treated dynamically is an N-body problem with complexity of $\mathcal{O}(\mathrm{N}^2)$. 
This is too computationally expensive to directly solve for the large number of photons required to statistically analyse the interaction.
Through the use of a tree code, as demonstrated by Jansen \textit{et al.} \cite{jansen2016}, the problem complexity can be reduced to $\mathcal{O}(\mathrm{N\, log\,N})$.
Here, by treating one source as a static field, the complexity is reduced further to $\mathcal{O}(\mathrm{N})$.
However, the type of photon-photon interactions which can be modelled are also constrained. 
By defining the static photon field with $n(\epsilon, \theta, \phi)$ neither spatial or temporal gradients are accounted for.
It is trivial to redefine the static field also as a function of position, $n(\mathbf{r}, \epsilon, \theta, \phi)$, and account for spatial gradients by constructing the full field through multiple sub-fields with different $n(\epsilon, \theta, \phi)$.
But it is not possible to relax the temporal gradient constraint due to the static nature of the Geant4 computational domain. 
Therefore, the module can only be used when one photon source is effectively stationary over the full duration of the interaction.

\begin{figure}
  \includegraphics[width=\linewidth]{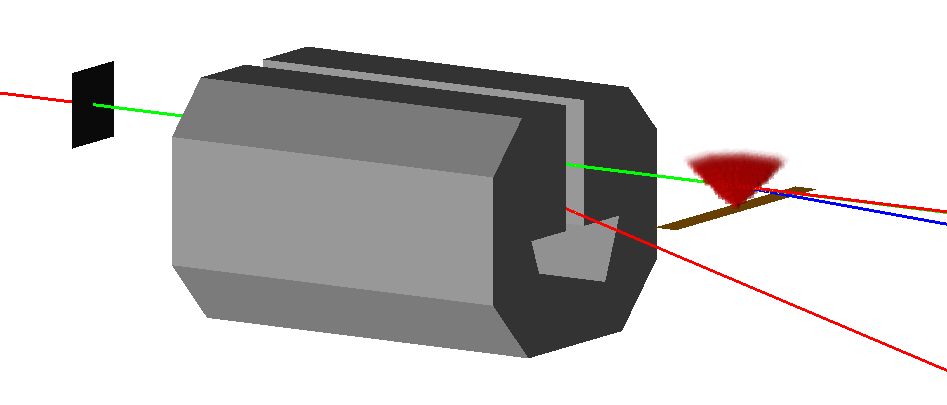}
  \caption{Geant4 visualisation of basic Breit-Wheeler experiment. An electron (red) produces a high energy gamma ray (green) in a bismuth target. The particles propagate through an on-axis magnet, diverting seed electron. The gamma ray propagates into an X-ray field annihilating and forming an  $e^+e^-$ pair.}
  \label{fig:geantPic}
\end{figure}

Although it constrains the type of experiment which can be modelled, developing the module within Geant4 allows us to make use of its extensive toolkit. 
It has the capability to model particle tracking, the geometry of experiments and the response of detectors to energetic particles.
Combining these capabilities with the module discussed here, start-to-end simulations can be performed. 
This includes: the generation of the gamma ray beam in the bremsstrahlung target; the background noise; the tracking of pairs through the analyser magnets and the particle detector interaction. 
This can be used for experimental parameter sensitivity analysis and optimisation. 
The ability to fully visualise the setup with Geant4 (see figure \ref{fig:geantPic}) also makes experimental design much quicker \cite{KettlearXiv2021}. 

\section{Module Demonstration}
Here we demonstrate how the module can be used to analyse a BW detection experiment similar to the Kettle \textit{et al.} experiment discussed in section \ref{sec:1} but with a simplified geometry and representation of the X-ray source. The setup is shown in figure \ref{fig:geantPic}. An electron beam is incident onto a $1\,\mathrm{mm}$ bismuth converter foil. 
This produces gamma rays through the bremsstrahlung process. 
The particles then pass through an on-axis magnet, removing the seed electrons and BH pairs generated in the converter foil. 
The gamma rays propagate into a $1\,$mm long X-ray field, generated by a laser heated germanium foil, and annihilate.

\begin{figure}[]
  \includegraphics[width=\linewidth]{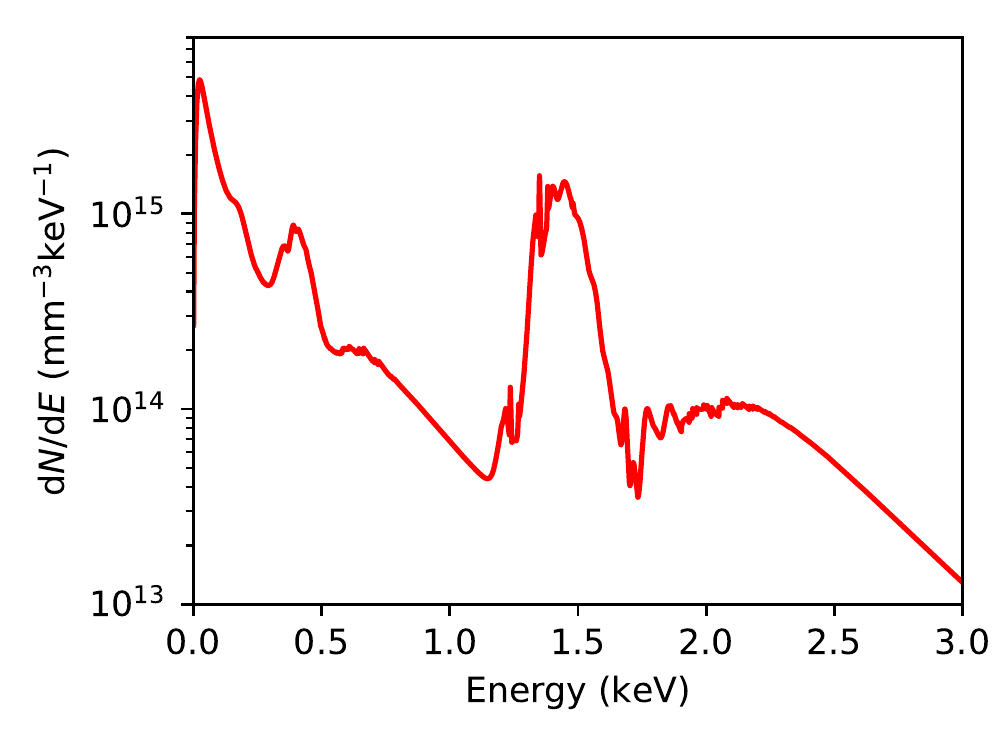}
  \caption{Radiation spectrum emitted from a static, solid density germanium plasma with an electron temperature of 350$\,$eV. This was generated using the atomic physics code Flychk \cite{flychk}.}
  \label{fig:fly}
\end{figure}

The X-ray field has spatial gradients, is anisotropic and has a non-blackbody energy spectrum, making this an ideal test for the module.
To model the spatial gradients, the field is split up into an array of $10\times 10 \times 10$ photon volumes, each with a different $n(\epsilon, \theta, \phi)$.
We assume that $n(\epsilon, \theta, \phi)$ is not correlated in energy and angle. This allows us to separate the energy and angular dependant parts, $n(\epsilon, \theta, \phi) = f(\epsilon)\Phi(\theta, \phi)$. 
$f(\epsilon)$ is then modelled using the atomic physics code Flychk \cite{flychk} and an example is shown in figure \ref{fig:fly} with a foil temperature of $350\, \mathrm{eV}$. 
$\Phi(\theta, \phi)$ is modelled using a Monte-Carlo X-ray photon tracking code. Here, photons are launched from an emitting plane and tracked through the photon volume array. By launching many photons, $\Phi(\theta, \phi)$ can be estimated.

\begin{figure} 
  \includegraphics[width=\linewidth]{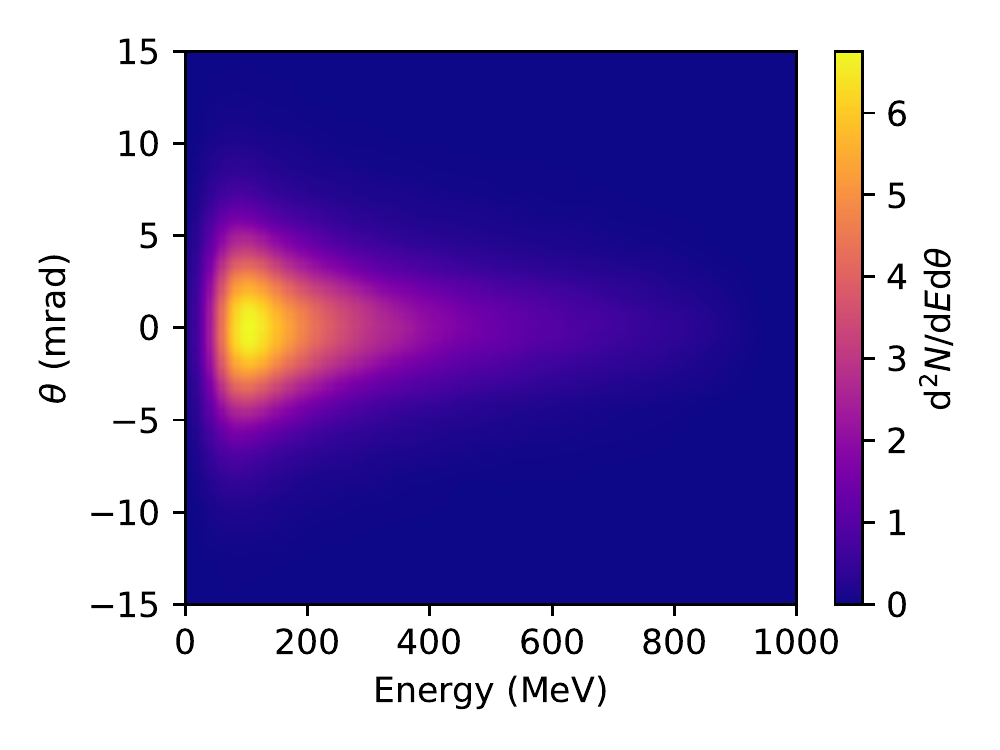}
  \caption{Positron angle and energy distribution.}
  \label{fig:posSpec}
\end{figure}

Figure \ref{fig:posSpec} shows the angle and energy probability distribution for BW positrons produced with a 1$\,$GeV electron beam and a burn-through foil electron temperature of 350$\,$eV.
From figure \ref{fig:posSpec} we can see the positron beam is well collimated with a divergence of $\sim10\,\mathrm{mrad}$.
The divergence of the positrons decreases with increasing energy.
The particle beaming is a result of the gamma ray carrying almost all the momentum of the interaction. 
This is a key advantage of the Pike \textit{et al.} detection scheme. 
If the positron beam had a large divergence, which is the case for two symmetric photon sources, to prevent a low detection efficiency the interaction region would have to be surrounded by detectors. 
A large number of high divergence background BH pairs would also be detected.
However, with a highly beamed source, an analyser chicane can be used to select positrons within a specific divergence and energy band and transport them to an area with low noise to be detected.

\begin{figure} []
  \includegraphics[width=\linewidth]{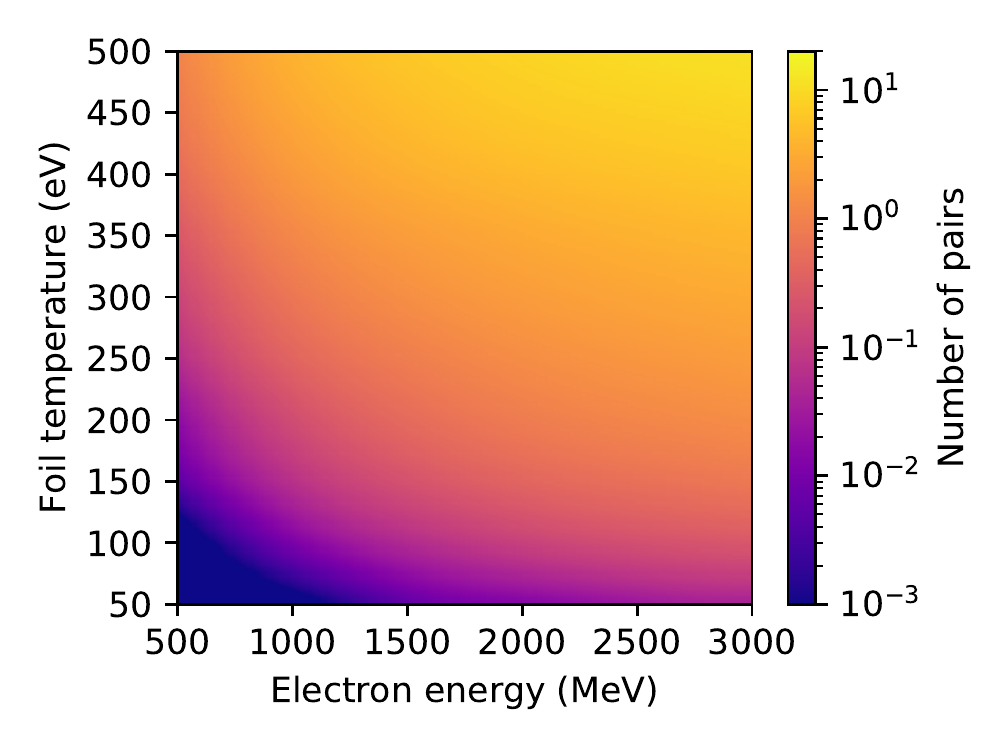}
  \caption{Breit-Wheeler pair yield against X-ray foil temperature and electron beam energy.}
  \label{fig:pairnumb}
\end{figure}

With the increased computational efficiency, as a result of implementing a GPR, the module can be used to perform scans over experimental parameters.
This is demonstrated in figure \ref{fig:pairnumb} where we can see how the yield of BW pairs is affected by the burn-through foil temperature and the LWFA electron beam energy.
Each simulation consisted of $10^9$ electrons (160$\,$pC) and the electron beam energy was varied from 500$\,$MeV to 3000$\,$MeV and foil temperature from $50\,$eV to $500\,$eV. 
The Gemini laser facility is capable of producing $2\,$GeV electron beams with a charge of $<100\,$pC.
Therefore, these results suggest that with a foil temperature of $>150\,$eV, it is possible to generate $>1$ BW pair per shot.
Utilising the relatively high repetition rate of the Gemini laser facility, over multiple shots a detectable number of BW pairs above background could be produced.

\section{Conclusion}
As the power of laser facilities has increased in recent years, it is now theoretically possible to produce a detectable number of BW pairs in the laboratory. 
However, the expected signal-to-noise ratio of such an experiment is low, making detailed numerical modelling vital. Here, we have presented the development of a new linear BW module for Geant4. 
This will allow signal-to-noise ratio calculations to be performed within a single framework.
We have shown how a Gaussian process regression can be used to greatly increase the rate at which photons are simulated without a loss in the accuracy of the module.
We have used this increase in efficiency to perform a parameter scan which suggests that $>1$ BW pair per shot can be produced using the Gemini laser facility.

The module presented here can be readily extended to include other particle-photon interactions such as Compton scattering ($e^- \gamma \rightarrow e^- \gamma$) and photon-photon scattering ($\gamma \gamma \rightarrow \gamma \gamma$). 
Photon-photon scattering from two real photons is another QED process with astrophysical relevance that has never been directly observed in the laboratory.
The difficulties of observing the process arise due to the relatively small cross-section ($\sim \alpha^2$ times smaller than $\sigma_{\mathrm{BW}}$) and the incoming and outgoing particles all being photons, making background subtraction challenging.
The scattering of quasi-real photons in the collision of heavy ion beams has recently been observed by the ATLAS detector at the Large Hadron Collider \cite{Collaboration2017}.
With improvements to the experimental scheme from section \ref{sec:1}, it may also be possible to study photon-photon scattering of two real photons, which could be modelled using this module.

\bibliography{BWGEANTmodel}

\end{document}